# The 5*f* localization/delocalization in square and hexagonal americium monolayers: A FP-LAPW electronic structure study

Da Gao and Asok K. Ray*
Physics Department
P. O. Box 19059
University of Texas at Arlington
Arlington, Texas 76019

*akr@uta.edu.

# Abstract

The electronic and geometrical properties of bulk americium and square and hexagonal americium monolayers have been studied with the full-potential linearized augmented plane wave (FP-LAPW) method. The effects of several common approximations are examined: (1) non-spin polarization (NSP) vs. spin polarization (SP); (2) scalar-relativity (no spin-orbit coupling (NSO)) vs. full-relativity (i.e., with spin-orbit (SO) coupling included); (3) local-density approximation (LDA) vs. generalized-gradient approximation (GGA). Our results indicate that both spin polarization and spin orbit coupling play important roles in determining the geometrical and electronic properties of americium bulk and monolayers. A compression of both americium square and hexagonal monolayers compared to the americium bulk is also observed. In general, the LDA is found to underestimate the equilibrium lattice constant and give a larger total energy compared to the GGA calculations. While spin orbit coupling shows a similar effect on both square and hexagonal monolayer calculations regardless of the model, GGA *versus* LDA, an unusual spin polarization effect on both square and hexagonal monolayers is found in the LDA results as compared with the GGA results. The 5*f* delocalization transition of americium is employed to explain our observed unusual spin polarization effect. In addition, our results at the LDA level of theory indicate a possible 5*f* delocalization could happen in the americium surface within the same Am II (fcc crystal structure) phase, unlike the usually reported americium 5*f* delocalization which is associated with crystal structure change. The similarities and dissimilarities between the properties of an Am monolayer and a Pu monolayer are discussed in detail.

# I. Introduction

Recent years have seen increased interests in the studies of strongly correlated and heavy fermion systems, including the actinides [1-6]. As is known, the actinides are characterized by a gradual filling of the 5f-electron shell with the degree of localization increasing with the atomic number Z along the last series of the periodic table. The open shell of the 5f electrons determines the magnetic and solid-state properties of the actinide elements and their compounds and understanding the quantum mechanics of the 5f electrons is the defining issue in the physics and chemistry of the actinide elements. These elements are also characterized by the increasing prominence of relativistic effects and their studies can, in fact, help us to understand the role of relativity throughout the periodic table. Narrower 5*f* bands near the Fermi level, compared to 4*d* and 5*d* bands in transition elements, is believed to be responsible for the exotic nature of actinides at ambient condition [6]. The 5f orbitals have properties intermediate between those of localized 4f and delocalized 3d orbitals and as such, the actinides constitute the "missing link" between the d transition elements and the lanthanides [1]. Thus a proper and accurate understanding of the actinides will help us understand the behavior of the lanthanides and transition metals as well.

Given the importance of 5f-electron contribution to binding [7], the actinides are typically divided in two groups: the lighter actinides, Th to Pu, are characterized by itinerant 5*f* electron behavior, implying that their 5*f* electrons do take part in bonding. The other group, Am and beyond, is named the heavy actinides and is characterized by localized 5*f* electron behavior. The nature of the 5*f*-electrons of the actinides is also illuminated by their fascinating crystal structure and atomic volume behavior [8]. The crystal structures of the light actinides become increasingly complex with the increase of the atomic number and plutonium has a monoclinic structure, with sixteen atoms per unit cell, in the ground state. But the heavy actinides typically favor high symmetry close packed structure, with americium in the double-hexagonal close-packed structure (dhcp) in the ground state. As regards the atomic volumes, the light actinides show a parabolic decreasing atomic volume from Th to Pu [9]. But from Pu to Am, this trend reverses with a 40% larger atomic volume found in Am compared with the atomic volume of Pu.

The pivotal position occupied by americium in the actinides, specifically in regards to the transition of the behavior of the 5f electrons from itinerant to delocalized, has attracted the interests of both theoreticians and experimentalists. Experimentally, the X-ray and high-resolution UV photoemission spectroscopy of the conduction band of Am concluded that the 5f electrons in Am are localized [10]. Recently, high pressure measurements of the resistivity of Am have been reported to 27 GPa and down to temperatures of 0.4K. An unusual dependence of superconducting temperature on pressure was deduced and at pressures of about 16 GPa the 5f electrons change from localized to itinerant and the crystal structure becomes complex [11]. The 5*f* delocalization of americium under pressure have indeed been thoroughly investigated both experimentally and theoretically with contradictory results [12-14]. It has been claimed that a critical structural link between americium under pressure and the preceding element plutonium is possible [13].

However, to the best of our knowledge, *no* study exists in the literature about the Am surface. Electronic structure studies of the Am surface is crucial not only from the point of view of a proper understanding of the Am metal in general but also for an understanding of the delocalization-localization transition, a matter of considerable controversy in studies of the actinides. Thus the primary motivation of this study is a first-principles electronic structure study of the Am surface. Such studies will also lead to a better understanding of the surface corrosion mechanisms which is not only scientifically challenging but also environmentally beneficial especially for actinides given their varying levels of toxicity. The present work uses an isolated Am monolayer to model the Am surface or an ultra-thin film. Although, in general, a monolayer is a rather poor approximation to the semi-infinite surface, one can deduce significantly useful information about bonding properties and the validity of commonly used theoretical approximations that is not readily available from thicker surface slabs calculations. In addition, study of the relaxation of an isolated monolayer compared with its bulk analog can provide knowledge of the stress that the remainder of the solid exerts on the outer layer. Thus it may provide guidance in selecting substrates to be used for epitaxial deposition of a single monolayer under laboratory conditions.

This study has thus focused on square and hexagonal Am monolayers that correspond to the (100) and (111) surfaces of Am II. We also studied bulk Am II for a direct comparison of bulk properties with the properties of the square and hexagonal monolayers. There are two main reasons for selecting Am II for such a study: First, this fcc structure has been experimentally determined for moderate pressures of Am [15] and the experimental data is readily available but controversial [16 – 18]. Second, Am II to Am III transition is attributed to 5$f$ electron delocalization [13]. Furthermore, it could be compared to our published monolayer results of δ-Pu with a fcc structure [19] and this might lead to a better understanding of the localization and/or delocalization of the Am 5f-electrons.

## II. Computational methods

A mixed basis APW + lo/LAPW method as implemented in the WIEN2K suite of softwares is employed in our calculations [20]. The addition of the new local orbital (lo) gives the radial basis functions more variational flexibility. This new scheme has successfully reduced the basis sets (up to 50%) and thus the corresponding computing time (up to an order of magnitude) [21]. On the other hand, the LAPW basis is energy-independent and is crucial for avoiding the nonlinear eigenvalue problem. So in the WIEN2k program, the basis set APW + lo is used inside the atomic spheres for the chemically important orbitals that are difficult to converge, whereas the LAPW basis set is used for others. A gradient corrected Perdew–Berke–Ernzerhof (PBE) functional (Generalized Gradient Approximation GGA) and the local density approximation (LDA) to density functional theory (DFT) are used respectively to describe the exchange and correlation effects in order to study and compare effects induced by GGA *vs.* LDA [22-23]. As far as relativistic effects are concerned, core states are treated fully relativistically and two levels of treatments are implemented for valence states: (1) a scalar relativistic scheme (without spin-orbit coupling) that describes the main contraction or expansion of various orbitals due to the mass-velocity correction and the Darwin s-shift [24] and (2) a fully relativistic scheme with spin-orbit coupling included in a second-variational treatment using the scalar-relativistic eigenfunctions as basis [25]. For the bulk calculations, a fcc unit cell with one atom is used. A constant muffin-tin radius ($R_{mt}$) of 1.95a.u and large plane-wave cut-off $K_{max}$ determined by $R_{mt}K_{max}=10.0$ are used for all calculations. The Brillouin zone is sampled on a uniform mesh with 104 irreducible K-points for the fcc bulk americium. The square and hexagonal monolayers of americium are modeled by a periodically repeated fcc Am surface slab with one Am layer separated by a 15 Å vacuum gap. Sixteen irreducible K-points have been used for reciprocal-space integrations in the surface calculations. For both bulk Am and monolayer calculations, the energy convergence criterion is set to be 0.01 mRy.

## III. Results and discussions

As mentioned in the Introduction, this study concentrates on Am monolayers in (100) and (111) symmetries. For the sake of comparison, as mentioned before, we have also carried out Am bulk calculations. The objective here, apart from studying an Am monolayer, is also to compare the effects of various approximations on the electronic structure properties of the monolayer: (1) scalar-relativity *vs.* full-relativity (*i.e.* with spin-orbit coupling included); (2) non-spin polarized *vs.* spin-polarized; and (3) LDA *vs.*

GGA. Thus eight levels of theory, namely non-spin-polarized-no-spin-orbit-coupling (NSP-NSO), non-spin-polarized-spin-orbit-coupling (NSP-SO), spin-polarized-no-spin-orbit-coupling (SP-NSO), spin-polarized-spin-orbit-coupling (SP-SO) calculations at both LDA and GGA levels of theory have been employed for all bulk and monolayer calculations. For the bulk calculations, the volume optimization feature in the WIEN2k package is used to yield a series of different total energies with varying bulk volumes until a total energy minimum is reached. Then we applied the Murnaghan equation of state [26]

$$E = BV/\beta(1/(\beta-1)(V_0/V)^\beta + 1) \qquad (1)$$

to fit the total energy curve and obtain the equilibrium lattice constant and the bulk modulus. We have listed these results together with some of the available theoretical and experimental results [7, 14, 27-30] in Table 1.

Our results show that the LDA calculations always give a smaller lattice constant compared with the corresponding GGA results at the same level of calculation. With spin polarization included, this difference becomes more apparent. At the same time, spin-orbit coupling is observed to play an important role in predicting the bulk properties regardless of the model used, LDA *versus* GGA. For bulk modulus predictions, the fluctuation in GGA results is found to be much smaller than those given by the LDA calculations, indicating that GGA is more reliable than LDA for such calculations. With both spin polarization and spin-orbit coupling included, our GGA calculated equilibrium lattice constant and bulk modulus are 9.32 a.u. and 51.52 GPa respectively, in excellent agreement with the experimental values of 9.26 a.u. and 45 GPa, the percent difference in the lattice constant being only 0.6 percent.

For the Am monolayers, all calculations are carried out at eight different theory levels as well. To find the optimized surface geometry structure, a series of total energy values are calculated by varying nearest-neighbor-distance (nnd) for a wide range of values from around 4.3 a.u. to around 7.3 a.u.. After finding the approximate position of the total energy minimum, a second similar calculation, *i.e.* total energy *versus* the nearest-neighbor distance, is carried out with a fixed 0.01 a.u. nearest-neighbor-distance step difference to obtain the final equilibrium nearest-neighbor-distance. The work function $W_f$ of both monolayers is also calculated at eight different theory levels according to:

$$W = V_0 - E_F \qquad (2)$$

where $V_0$ is the Coulomb potential energy at half the height of the surface slab including the vacuum layer and $E_f$ is the Fermi energy. We also calculated the spin magnet moment for both bulk and monolayer calculations. The results are listed in Table 2 and plotted in Figs. 1– 4 respectively.

The previously observed trend that LDA always underestimates the nearest-neighbor distances of bulk Am compared with the corresponding GGA calculation still exists for the square and hexagonal monolayers. The spin-orbit coupling is found to have a small effect on the nearest-neighbor distance calculated in LDA, varying from 1.3 percent to 2.3 percent for both square monolayer and hexagonal monolayers while for the corresponding GGA calculation, this spin-orbit coupling effect is increased to the range of 1.7 percent to 4.4 percent for both monolayers. This confirms that for the GGA model, spin-orbit coupling effects must be included to achieve a better approximation [19]. A general compression of both monolayers is also observed as we compare the equilibrium nearest-neighbor distances with those values of the bulk americium. For the square monolayer at the LDA level, the compression varies from 17.45 to 31.25 percent and at the GGA level, the corresponding values are 5.01 to 21.34 percent. At the LDA level for the hexagonal monolayer, the compression varies from 12.95 to 27.50 percent and at the GGA level, the corresponding values are zero to 17.38 percent. This indicates that the atoms in the outer layers of the (111) and (100) surfaces of Am would compress if it were not for a large opposing stress exerted by the remainder of the solid. Furthermore, the current results suggest that it should be possible to grow a thin film of Am on a substrate with a rather smaller constant than that of bulk Am. The same conclusions were found to be true for our previous study on Pu monolayers, using the linear combination of Gaussian type orbitals – fitting function (LCGTO-FF) method [19]. For the Pu square monolayer, the compression varied from 17.8 to 21.0 percent and for the hexagonal monolayer, the compression varied from 12.0 to 15.1 percent, all depending on the level of theory employed.

The spin polarization, on the other hand, has no observable effect on the nearest-neighbor distance in LDA calculations for both square and hexagonal monolayers whereas a significant spin polarization effect is observed in GGA calculations for both monolayers. To be specific, without spin-orbit coupling, the spin polarization could affect the predicted nearest-neighbor distance value up to around 60 percent, and with spin-polarization included with spin-orbit coupling, this value drops to about 37 percent. In general, the GGA calculation with both spin polarization and spin-orbit coupling included gives more accurate predictions. This is in agreement with our bulk calculations.

The observed spin polarization effect difference between LDA and GGA mentioned above can be understood from two aspects: First, LDA underestimates the equilibrium nearest-neighbor distance and thus enhances hybridization and inhibits spin polarization. Second, it is well known that the 5$f$ electrons in Am is localized under normal conditions and GGA favors density inhomogeneity more than LDA [22]. This is further confirmed by the graphs shown in Fig.1 and Fig.2. The LDA spin polarization results of square monolayer (Fig. 1) and hexagonal monolayer (Fig. 2) specially featured two energy minima with varying nearest-neighbor distances. The underlying physics to explain such a feature is that 5$f$ Am delocalization is induced as it is compressed [12 – 14]. To confirm this observed 5$f$ delocalization, we explore the $f$ –band density of states

(DOS) of the LDA (SP-NSO) calculation as an example at both minima, namely the nearest-neighbor distance of 4.64 a.u. and 6.19 a.u. respectively, for the hexagonal monolayer and the corresponding $f$-band DOS graph is shown in Fig. 5 and Fig. 6. It is apparent that the 5$f$ electrons at the first minimum are already delocalized with the highest peak beyond the Fermi energy whereas at the second minimum the 5$f$ electrons are still localized with the highest spin up peak well below the Fermi energy. Same conclusions prevail at the other levels of theory.

The spin magnetic moment as a function of the nearest-neighbor-distance for the square monolayer and the hexagonal monolayer in the GGA and LDA calculations are plotted in Figs. 3 and Fig. 4 respectively. The LDA calculations are found to consistently predict a smaller spin magnetic moment compared with the corresponding GGA calculations, which is in general agreement with the results obtained for Pu monolayers [19]. In order to further understand the behavior of the spin magnetic moments of the Am monolayers, we have examined the $f$-band density of states ($f$-DOS) of the hexagonal monolayer at three representative points, namely two minimum points of LDA (SP-NSO) and the minimum point of GGA (SP-NSO) calculations (Figs. 5 – 7). Fig. 5 shows that at the first minimum energy point of LDA there is no observable difference between up and down electrons in the monolayer leading to a zero spin magnetic moment while at the second minimum energy point of LDA there is a significant observable difference between up and down electrons in the monolayer (Fig. 6), which could be easily judged by the height difference of up and down spins, leading to a significant spin magnetic moment. On the other hand, an even more observable difference between up and down electrons in the GGA hexagonal monolayer is noted in Fig. 7, as indicated by the fact that a much higher second spin up peak is found in Fig.7 but not in Fig. 6, causing a higher spin magnetic moment in GGA than that in the corresponding LDA.

According to the results presented and the fact that LDA behaves significantly different for the Am monolayer, we propose that LDA calculations might be used as a tool to signal the 5*f* delocalization. At one LDA theory level (SP-NSO), we have examined the total energy as a function of the nearest-neighbor distance of δ-Pu square monolayer and compared it with the corresponding Am square monolayer result. The comparison of the result between Am and δ-Pu is plotted in Fig.8. It is possible that the 5*f* electrons in a highly compressed monolayer of δ-Pu are delocalized and very different from the localized 5*f* electrons of Am. As a result, a sharp peak is observed for Am but not for δ-Pu and indicates that the 5*f* electrons of Am have changed from localized to the delocalized state, which is featured in the *f*-DOS discussed above as well.

Finally, the work functions for the square monolayer of Am vary over a range of 2.77eV to 4.37eV, whereas for the hexagonal monolayer, the range is from 2.85eV to 4.80eV. The GGA values are always smaller than the corresponding LDA values, with the GGA (SP-NSO) values being the smallest for both monolayers. Using the different LCGTO-FF DFT methodology, we have previously found that the work functions for a square monolayer of Pu varied from 4.36 to 4.92eV, and the values for a hexagonal monolayer varied from 4.28eV to 4.85eV. For the Pu monolayer, the GGA values at the NSP-SO level provided the smallest value. For the majority of the cases, the work functions for the Am monolayer are lower than the corresponding values of the Am monolayer indicating that it would require more energy to "ionize" a Pu monolayer compared to an Am monolayer. These comments should apply irrespective of the computational methodologies used.

## IV. Conclusions

We have studied the electronic and geometrical properties of square and hexagonal americium monolayers via the full-potential all-electron density functional calculations at eight different theory levels. The bulk properties are also investigated and have been compared to the monolayer results. A mixed basis APW +lo/LAPW embedded in the WIEN2k software has been employed for our calculations. The effects of several common approximations have been examined: (1) non-spin polarization (NSP) vs. spin polarization (SP); (2) scalar-relativity (no spin-orbit coupling (NSO)) vs. full-relativity (i.e., with spin-orbit (SO) coupling included); (3) local-density approximation (LDA) vs. generalized-gradient approximation (GGA). Our results indicate that both spin polarization and spin orbit coupling play important roles in determining the geometrical and electronic properties of americium bulk and monolayers.

The 5*f* delocalization transition of Am II is found as the americium monolayers are compressed, unlike the reported americium 5*f* delocalization which is usually associated with crystal structure changes [12 – 14]. We also propose that the LDA calculation might be applied as a tool to detect the 5*f* localized-delocalized transition.

# Acknowledgments

This work is supported by the Chemical Sciences, Geosciences and Biosciences Division, Office of Basic Energy Sciences, Office of Science, U. S. Department of Energy (Grant No. DE-FG02-03ER15409) and the Welch Foundation, Houston, Texas (Grant No. Y-1525).

Table 1. Calculated bulk Am equilibrium lattice constant $a_0$ (in a.u.) and the corresponding bulk modulus B (in GPa).

| Method | $a_0$ (a.u.) | B (GPa) |
|---|---|---|
| LDA (NSP-NSO) [a] | 7.54 | 297.66 |
| LDA (NSP-SO) [a] | 8.14 | 323.58 |
| LDA (SP-NSO) [a] | 9.05 | 27.62 |
| LDA (SP-SO) [a] | 8.26 | 138.27 |
| GGA (NSP-NSO) [a] | 7.73 | 170.74 |
| GGA (NSP-SO) [a] | 8.22 | 251.74 |
| GGA (SP-NSO) [a] | 9.87 | 13.89 |
| GGA (SP-SO) [a] | 9.32 | 51.52 |
| LMTO (LDA-NSP-NSO) [b] | 7.47 | N/A |
| LMTO (LDA-SP-NSO) [b] | 9.08 | 45 |
| DLM (LDA-SP-NSO) [c] | 9.11 | 43 |
| GGA + OP [d] | 8.78 | 43 |
| FPLAPW (GGA-NSP-SO) [e] | 8.04 | N/A |
| FPLAPW (GGA-SP-SO) [e] | 9.51 | N/A |
| FPLAPW (GGA-AFM-SO) [e] | 9.11 | N/A |
| Exp [d] [f] [g] | 9.26 | 29.4, 45 |

(a) The present work; (b) Ref. 7; (c) Ref. 27; (d) Ref. 29; (e) Ref. 14; (f) Ref. 29; (g) Ref. 30.

Table 2. A comparison of the bulk Am, square monolayer, and hexagonal monolayers.

| System | Theory | nnd (a.u.) | MM (μB) | $W_f$ (eV) |
|---|---|---|---|---|
| Square Monolayer | LDA (NSP-NSO) | 4.40 | | 4.36 |
| | LDA (NSP-SO) | 4.50 | | 4.23 |
| | LDA (SP-NSO) | 4.40 | 0.00 | 4.37 |
| | LDA (SP-SO) | 4.47 | 0.00 | 4.23 |
| | GGA (NSP-NSO) | 4.46 | | 4.02 |
| | GGA (NSP-SO) | 4.57 | | 3.90 |
| | GGA (SP-NSO) | 6.55 | 7.80 | 2.77 |
| | GGA (SP-SO) | 6.26 | 7.32 | 3.01 |
| Hexagonal Monolayer | LDA (NSP-NSO) | 4.64 | | 4.78 |
| | LDA (NSP-SO) | 4.70 | | 4.79 |
| | LDA (SP-NSO) | 4.64 | 0.00 | 4.79 |
| | LDA (SP-SO) | 4.70 | 0.00 | 4.80 |
| | GGA (NSP-NSO) | 4.72 | | 4.41 |
| | GGA (NSP-SO) | 4.80 | | 4.43 |
| | GGA (SP-NSO) | 6.72 | 7.64 | 2.85 |
| | GGA (SP-SO) | 6.59 | 7.44 | 2.99 |
| Am Metal | LDA (NSP-NSO) | 5.33 | | |
| | LDA (NSP-SO) | 5.76 | | |
| | LDA (SP-NSO) | 6.40 | 6.89 | |
| | LDA (SP-SO) | 5.84 | 5.32 | |
| | GGA (NSP-NSO) | 5.47 | | |
| | GGA (NSP-SO) | 5.81 | | |
| | GGA (SP-NSO) | 6.98 | 7.35 | |
| | GGA (SP-SO) | 6.59 | 6.88 | |

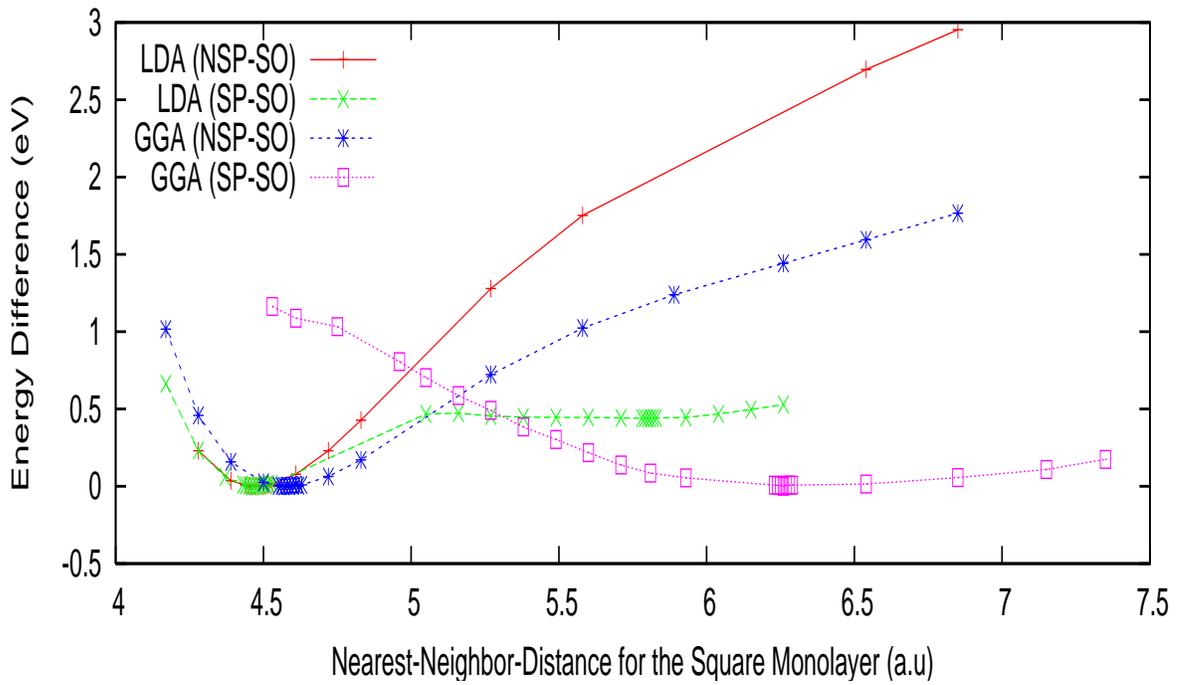

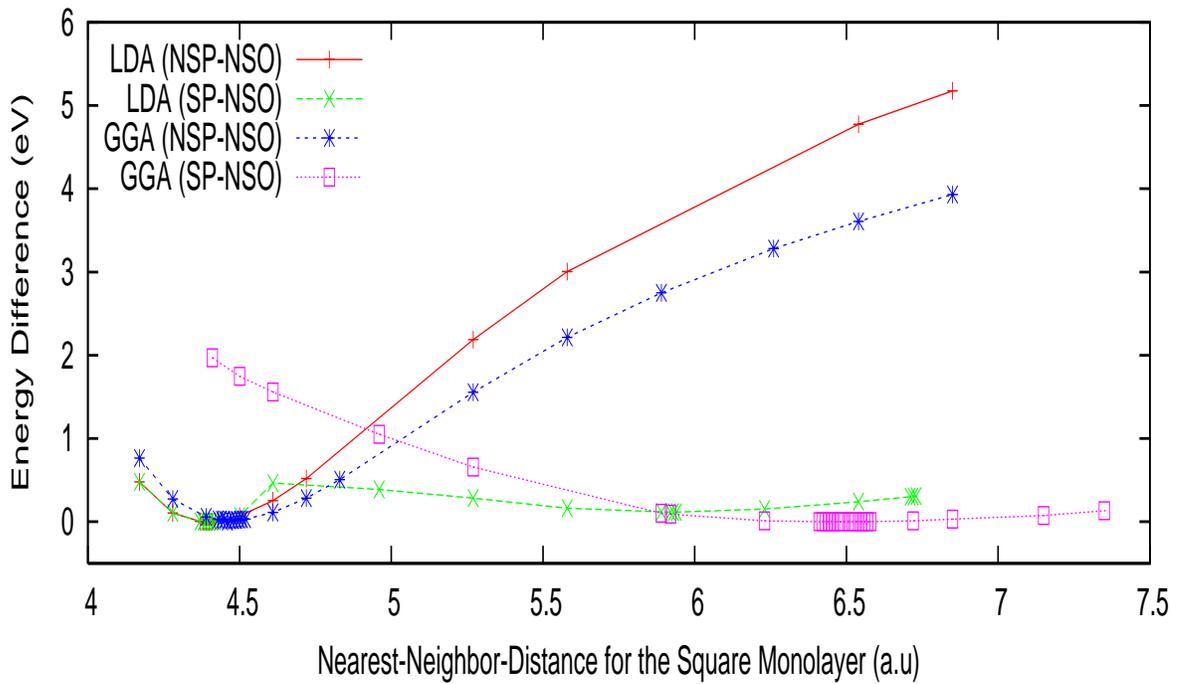

Fig.1 The energy difference $\Delta E = E_i - E_{min}$ versus the nearest-neighbor-distance (nnd) for the square monolayer, where $E_i$ is the total energy calculated with one nearest-neighbor-distance and $E_{min}$ is the minimum total energy obtained in that theory level's calculation

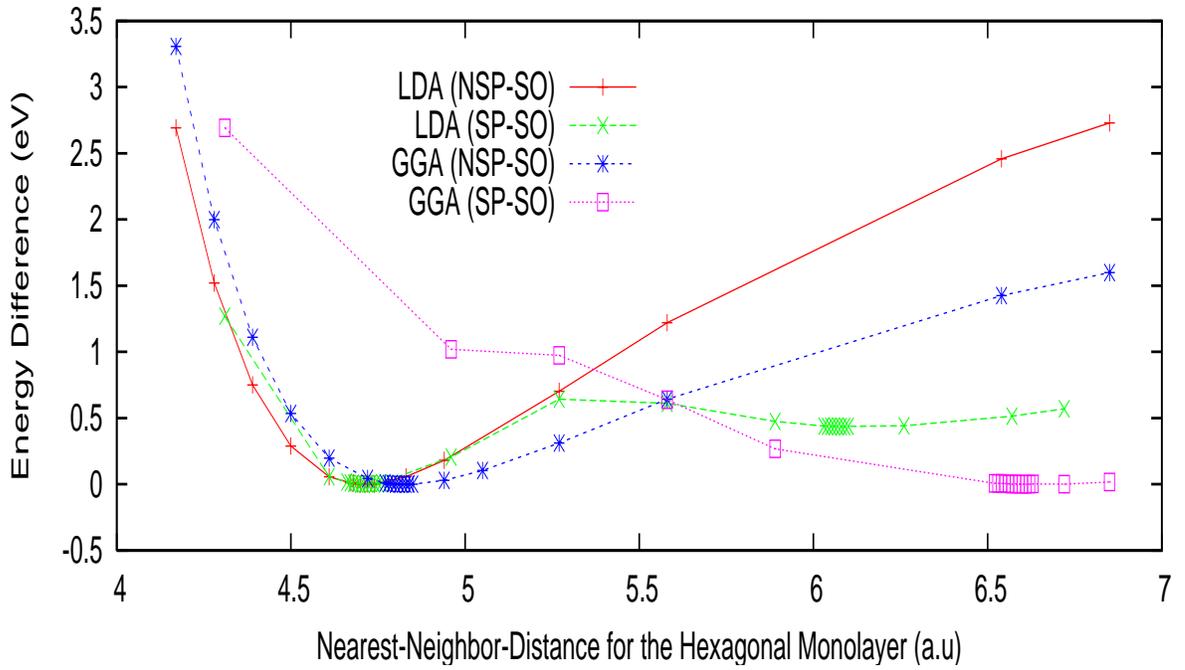

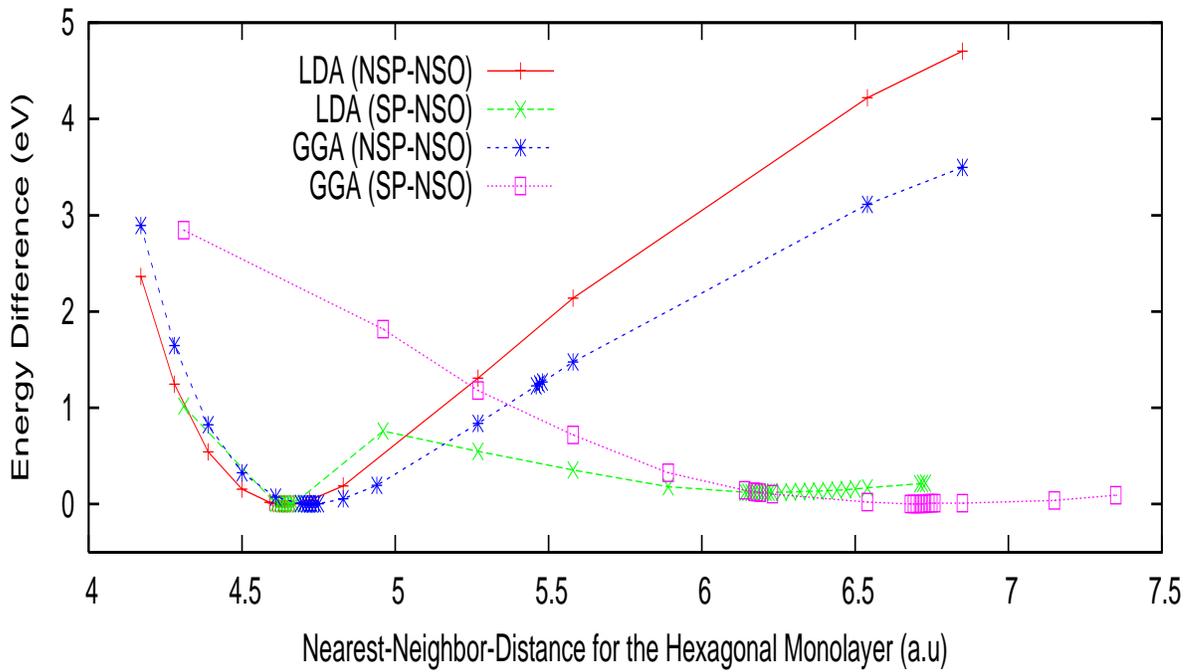

Fig. 2 The energy difference $\Delta E = E_i - E_{min}$ versus the nearest-neighbor-distance (nnd) for the hexagonal monolayer, where $E_i$ is the total energy calculated with one nearest-neighbor-distance and $E_{min}$ is the minimum total energy obtained in that theory level's calculation

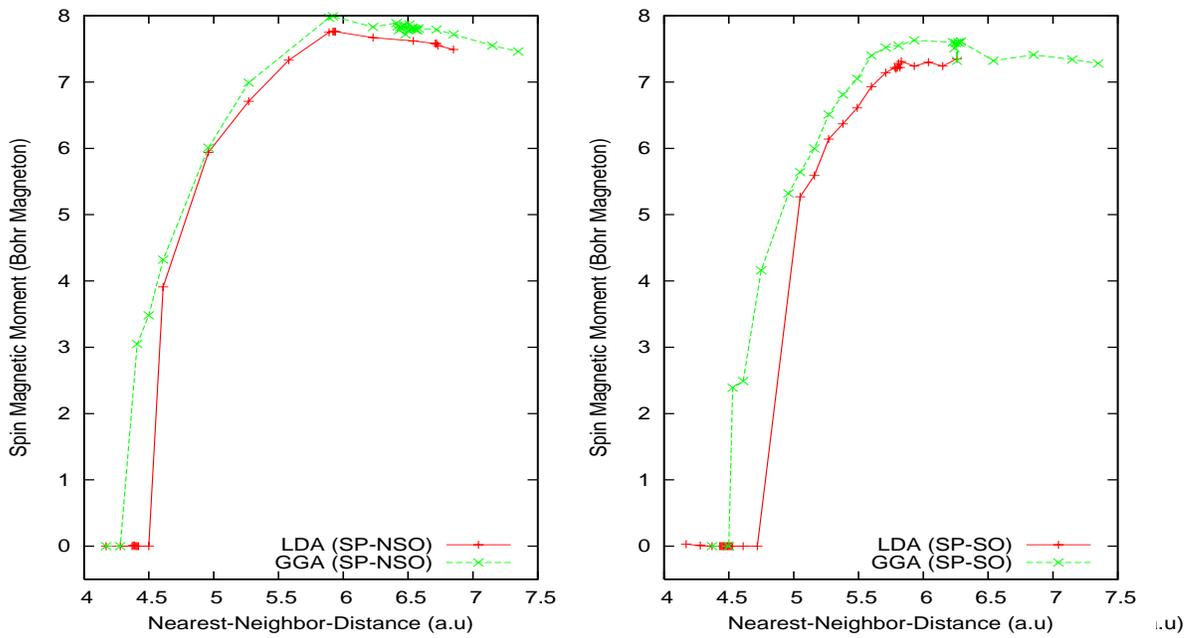

Fig.3 Spin magnetic moment *vs*. nearest-neighbor-distance (nnd) for the square monolayer in the GGA and LDA calculations.

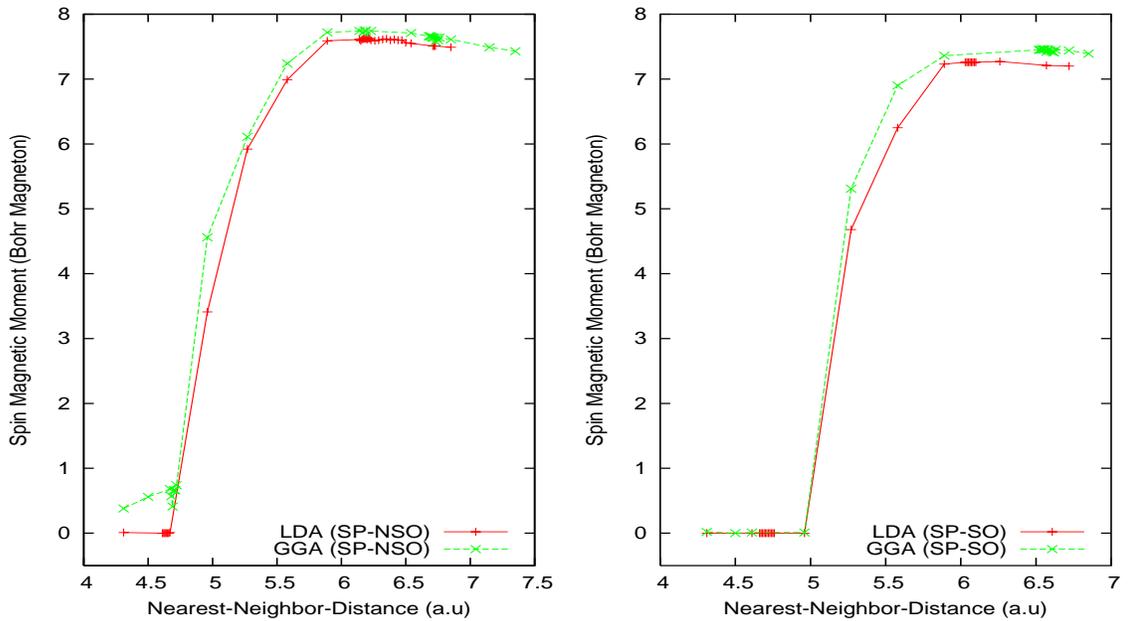

Fig. 4 Spin magnetic moment *vs*. nearest-neighbor-distance (nnd) for the hexagonal monolayer in the GGA and LDA calculations.

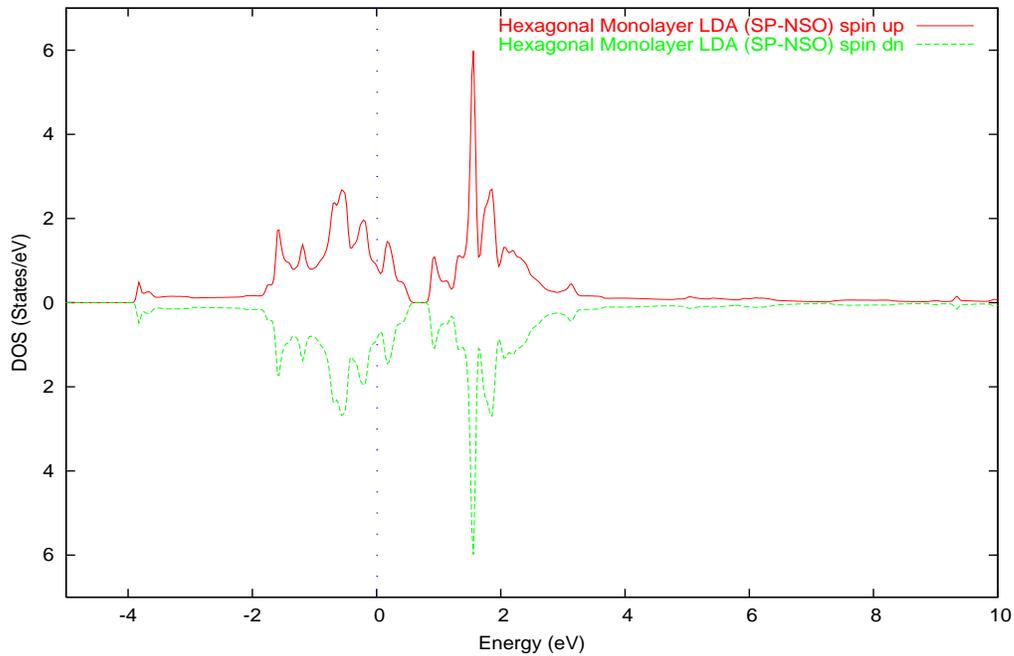

Fig.5. The partial density of states of spin-up and spin-down f bands for the hexagonal monolayer using LDA (SP-NSO) with the nearest-neighbor distance (nnd) equals 4.64 a.u.. Fermi energy is set at zero.

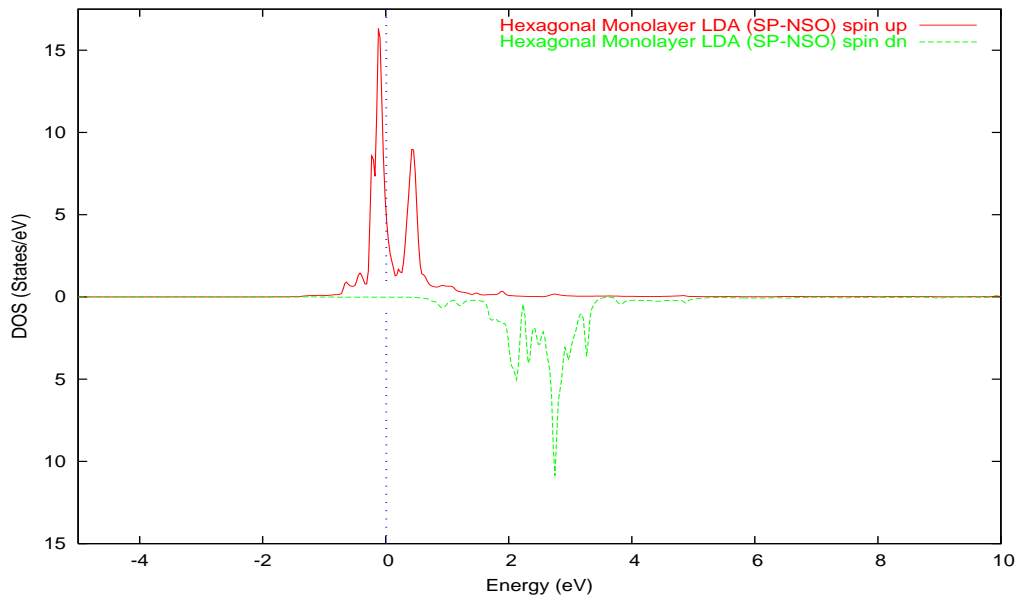

Fig. 6 The partial density of states of spin-up and spin-down f bands for the hexagonal monolayer using LDA (SP-NSO) with the nearest-neighbor distance (nnd) equals 6.19 a.u.. Fermi energy is set at zero.

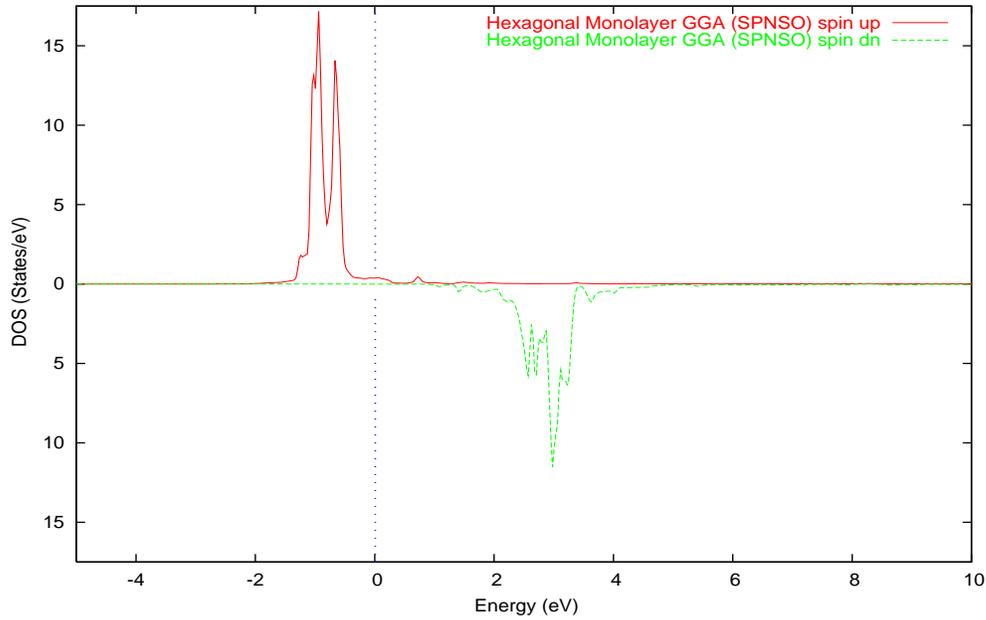

Fig. 7. The partial density of states of spin-up and spin-down f bands for the hexagonal monolayer using GGA (SP-NSO) with the nearest-neighbor distance (nnd) equals 6.72 a.u..Fermi energy is set at zero.

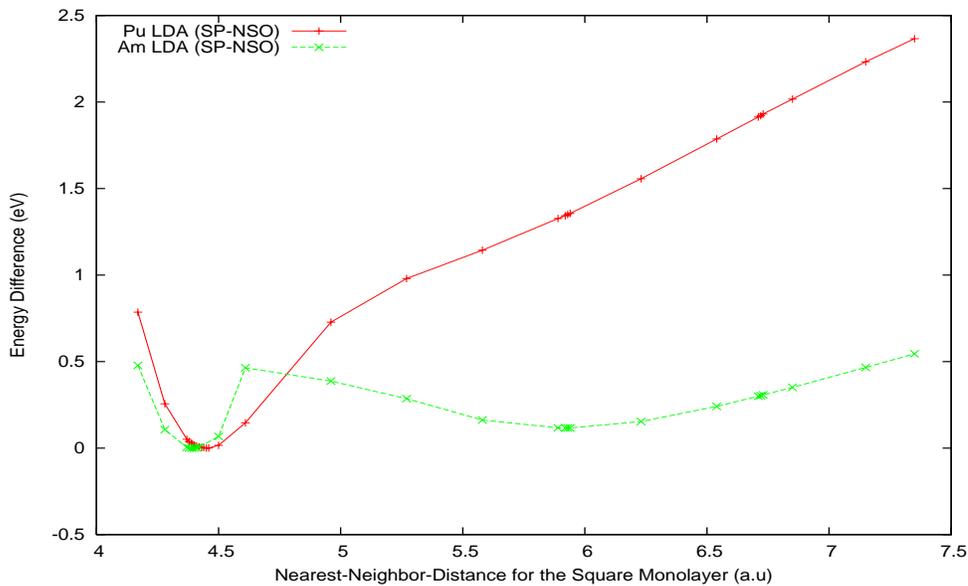

Fig.8. Comparison of the δ-Pu and Am II square monolayer energy difference $\Delta E = E_i - E_{min}$ results as a function of nearest-neighbor-distances (nnd) using LDA (SP-NSO), where $E_i$ is the total energy calculated with one nearest-neighbor-distance (nnd) and $E_{min}$ is the minimum total energy obtained in that theory level's calculation